\documentclass[conference]{IEEEtran}

\usepackage{ifpdf}

\usepackage{cite}

\usepackage{graphicx}
\usepackage{epstopdf} 
\usepackage{xcolor}
\ifCLASSINFOpdf

\fi

\usepackage[cmex10]{amsmath}
\usepackage{amsfonts,amssymb}

\usepackage{algorithmic}

\usepackage{array}

\ifCLASSOPTIONcompsoc
\usepackage[caption=false,font=normalsize,labelfont=sf,textfont=sf]{subfig}
\else
\usepackage[caption=false,font=footnotesize]{subfig}
\fi

\usepackage{dblfloatfix}

\usepackage{url}

\usepackage{arydshln}
\usepackage{enumerate}

\usepackage[mathcal]{euscript}
\newtheorem{theorem}{Theorem}
\newtheorem{lemma}{Lemma}

\newcommand{\sfm}[1]{\mathsf{#1}}

\def\beq{\begin{equation}}
\def\eeq{\end{equation}}

\usepackage{fancyhdr}

\fancypagestyle{firstpage}{
	\fancyhf{}
	\fancyhead[C]{Accepted for publication in GC'17 Workshops - LSASLUB}
}

\graphicspath {{figures/}}
\IEEEoverridecommandlockouts

\begin{document}
%
\title{Optimal Base Station Design with Limited Fronthaul: Massive Bandwidth or Massive MIMO?}

%
%
%


\author{\IEEEauthorblockN{Kamil Senel, Emil~Bj\"{o}rnson and Erik~G.~Larsson}
	\IEEEauthorblockA{Dept. of Electrical Engineering, Link\"{o}ping University, Link\"{o}ping, Sweden\\
		Email: \{kamil.senel, emil.bjornson, erik.g.larsson\}@liu.se}
	\thanks{This work was supported by ELLIIT, by the Swedish Research Council (VR), project 2015-05573 and by Swedish Foundation for Strategic Research.}}

\maketitle

\begin{abstract}
To reach a cost-efficient 5G architecture, the use of remote radio heads connected through a fronthaul to baseband controllers is a promising solution. However, the fronthaul links must support high bit rates as 5G networks are projected to use wide bandwidths and many antennas. Upgrading all of the existing fronthaul connections would be cumbersome, while replacing the remote radio head and upgrading the software in the baseband controllers is relatively simple. In this paper, we consider the uplink and seek the answer to the question: If we have a fixed fronthaul capacity and can deploy any technology in the remote radio head, what is the optimal technology? In particular, we optimize the number of antennas, quantization bits and bandwidth to maximize the sum rate under a fronthaul capacity constraint. The analytical results suggest that operating with many antennas equipped with low-resolution analog-to-digital converters, while the interplay between number of antennas and bandwidth depends on various parameters. The numerical analysis provides further insights into the design of communication systems with limited fronthaul capacity.   
\end{abstract}


\IEEEpeerreviewmaketitle

\section{Introduction}\label{sec:Introduction}

The fifth generation (5G) wireless networks are expected to not only support higher data rates, but also provide a uniform quality of service (QoS) throughout the network \cite{qual1000}. This is fundamentally a techno-economic problem of finding efficient technical solutions under cost and energy consumption constraints \cite{zander2017beyondUDB}.

The candidate technologies for satisfying the 5G demands for higher data rates are small-cell networks \cite{hoydis2011smallcell},\cite{hwang2013holisticSmallcell}, Massive MIMO \cite{erik2014MIMOmag}, and using large bandwidth in mmWave bands \cite{niu2015mmWave}. Small cell networks are based on ultra-dense deployment of low-cost, low-power base stations (BSs) and that achieve a higher signal-to-noise ratio by reducing the distance between the transmitter and receiver. However, it is challenging to control the inter-user interference in small cell networks \cite{andrews2016limits}. 
Equipping BSs with a large number of antennas is the main idea of Massive MIMO and this allows coherent beamformed transmission and spatial multiplexing of many users that enhances the spectral efficiency without relying on BS densification \cite{redBook}, but a cost-efficient deployment of  Massive MIMO requires successful utilization of low-cost hardware at BSs. The mmWave approach relies on utilization of idle spectrum in the range of $30$--$300$ GHz. However, a communication system that utilizes mmWave spectrum must overcome the hostile propagation characteristics exhibited by these high frequencies \cite{andrews2014whatWill5Gbe}.

\begin{figure}[tb]
	\begin{center}
		\includegraphics[trim=0cm 0cm 0cm 0cm,clip=true, scale = .5]{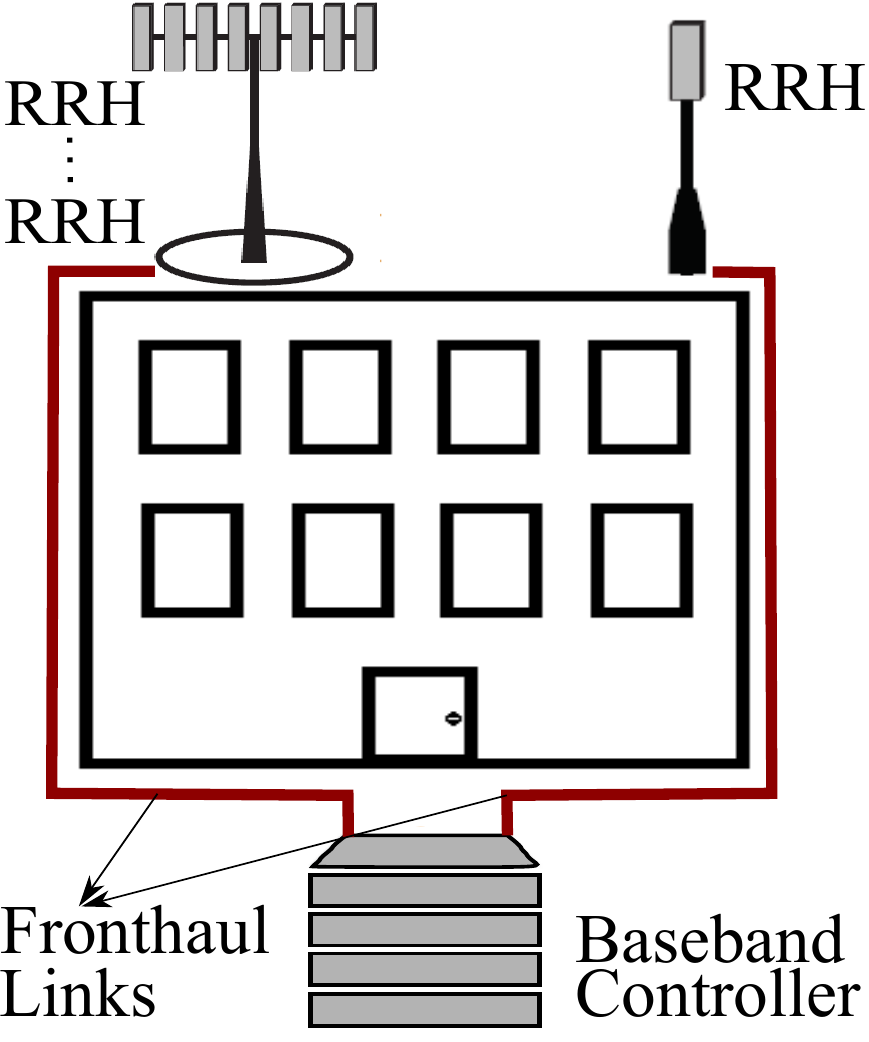}
		\caption{A network setup with RRHs connecting to BBU via fronthaul links.}
		\label{fig:FronthaulEx} \vspace{-8mm}
	\end{center}
\end{figure}

In traditional networks, baseband units (BBUs) and radio units are co-located. However, this setup suffers from several limitations such as underutilized dedicated resources, increased cost and energy 
consumption and limited flexibility \cite{5gppp2016RRHwhite}. These limitations can be addressed by detaching remote radio heads (RRH) and BBUs while maintaining the connection through fronthaul links as illustrated in Fig.~\ref{fig:FronthaulEx}.  This decoupling enables efficient resource utilization through network function virtualization. However, the connection between RRHs and BBUs must support high bit rates in order to be considered as a plausible 5G solution \cite{peng2015NFV}. The information exchanged between RRHs and BBUs usually consists of sampled radio signals which may require a rate up to $5$ Gbit/s for a signal with $100$ MHz bandwidth with 16-bit samples \cite{gomes2015fronthaul}. 
The required fronthaul rate grows linearly with the number of antennas and the bandwidth, which is a critical issue for Massive MIMO and mmWave deployment. Upgrading the existing fronthaul connections is costly, while replacing the RRHs and upgrading the software in the BBUs is relatively simple.

The use of low resolution analog-to-digital converters (ADCs) in MIMO systems has recently attracted considerable attention as low bit ADCs greatly reduces the implementation costs. For a wide range of signal to noise ratio (SNR) levels, low bit ADCs has been shown to perform well in various works \cite{chris2016uplinkOnebit,desset2015validation}. In \cite{jacobsson2017lowADCmimo}, an uplink MIMO system is considered and it is shown that the achievable rate for the unquantized system can be approached with ADCs with only a few bits of resolution.
However, the aforementioned papers consider a fixed bandwidth and neglect the fronthaul.



The main contribution of this work is to develop an analytical framework for designing the base station technology with limited fronthaul capacity and to provide guidelines for finding the operating characteristics that maximizes the sum rate. Furthermore, our work  highlights many fundamental tradeoffs that exist in practical communication systems.

%
%
%
%

\section{System Setup}\label{sec:SystemSetup}
We consider the uplink communication between a BS with $M$ antennas and $K$ single-antenna users. A total bandwidth of $B_w$ is available for the system and all of the signals are assumed to be uniformly sampled at Nyquist rate. An analog matched filter is used at the receiver and the quadrature and in-phase components of the signal at its output are separately sampled by identical ADCs with $b$-bit resolution.
 
 
Let $x_k[n]$ denote the normalized transmit signal from user $k$ and $P_k$ be the transmit power per sample of user $k$. 
The received signal at BS antenna $m$ at symbol time $n$ is   
\begin{equation}\label{eq:ReceivedSignalAntenna_m}
y_m[n] = \sum\limits_{k=1}^K \sum\limits_{l=0}^{L-1}\sqrt{P_k}g_{mk}[l]x_k[n-l] + z_m[n],
\end{equation}
where $z_m[n] \sim \mathit{CN}(0, N_0) $ represents the thermal noise at the receiver and it is modeled as a white stochastic process. The channel between user $k$ and antenna $m$ is described by an $L$-tap impulse response, $g_{mk}[l] = \sqrt{\beta_k} h_{mk}[l]$
where $\beta_k$ and $h_{mk}$ denotes the large and small scale fading components, respectively. 
The large scale fading components are assumed to be known at the BS, but the small scale fading is to be estimated. The mean value of the small scale fading is assumed to have zero mean with variance 
\begin{equation}\label{eq:PowerDelayProfile}
\sigma^2_k[l] = \mathbb{E}[|h_{mk}[l]|^2]
\end{equation}         
and the power delay profile is normalized as
\begin{equation}
\sum\limits_{l=0}^{L-1} \sigma^2_k[l] = 1.
\end{equation}  

The transmission is carried out via blocks of $N$ symbols which contain a cyclic prefix. Let $\sfm{h}_{mk}[v]$ denote the frequency response of the channel, then the relation between the input and output in the frequency domain is given by 
\begin{eqnarray} \label{eq:ReceivedSignalFreq-1}
\mathsf{y}_m[v] &=& \frac{1}{\sqrt{N}} \sum\limits_{n=0}^{N-1} y_m[n]e^{-j2\pi nv/N}\nonumber \\ &=& \sum\limits_{k=1}^K \sqrt{\beta_k P_k} \sfm{h}_{mk}[v]\sfm{x}_k[v] + \sfm{z}_m[v],
\end{eqnarray}
where $\sfm{x}_k[v]$ and $\sfm{z}_m[v]$ are the Fourier transform of the transmit signal and noise. 

The in-phase and quadrature components of the received signal are uniformly quantized by identical ADCs with $b$-bit resolution.
In practice, automatic gain control (AGC) amplifiers are utilized to adjust the input to the ADCs, to 
minimize the overload distortion and efficiently use the dynamic range. 

For a real ADC input $x[n]$, the quantized output is given by 
\begin{equation}\label{eq:Quantization}
x^q[n] = x[n] + q[n],
\end{equation}
where $q[n]$ represents the distortion due to quantization. Under the pseudo-quantization noise model (PQN), the distortion can be approximated as a uniformly distributed random variable which is uncorrelated with the input and noise \cite{sarajlic2016eeMIMOquantization}. The variance of $q$ is modeled as
\begin{equation}\label{eq:QuantizationMSE}
\mathbb{E}[|q|^2] \approx \frac{1}{3}X_{int}^2 2^{-2b} = E
\end{equation}  
where $[-X_{int}, X_{int}]$ defines the quantization interval. Even for low-bit resolutions, the model is shown to achieve an accurate representation of the quantization MSE \cite{sarajlic2016eeMIMOquantization}. 


\subsection{Uplink Data Transmission}
In this section, the performance of the uplink transmission is investigated. First a low-complexity channel estimation method is introduced. Then, a lower bound on the achievable rate during data transmission is derived. We consider a block fading model where the channel is constant in a block of $N$ channel uses, which contains both channel estimation and data transmission. The succeeding analysis is similar to the one provided in \cite{chris2016uplinkMultibit} which utilizes an empirical quantization distortion model whereas this work is based on \eqref{eq:Quantization}.      

Orthogonal pilot sequences of length $N_p$ are utilized for channel estimation. User $k$ transmits the pilot signal $\phi_k[n]$ defined by
\begin{equation} 
\sum\limits_{n=0}^{N_p-1}
\phi_k[n]\phi^*_i[n+l] =
\begin{cases}
\label{eq:PilotSignals}
N_p, &\text{if}~k = i, l=0 \\
0, &\text{if}~k \neq i, l\neq 0.
\end{cases}  
\end{equation}
The length of the pilot sequences must satisfy $N_p \geq KL$ in order to satisfy \eqref{eq:PilotSignals}. 
The resulting signal after correlation and quantization is 
\begin{eqnarray}\label{eq:CorrelatedSignal}
r_{mk}[l] &=& \frac{1}{\sqrt{N_p}}\sum\limits_{n=0}^{N_p - 1}y_m^q[n]\phi^*[n+l] \\ 
 &=& \sqrt{\beta_k P_k N_p \mu_m} h_{mk}[l] + q'_{mk}[l] + z'_{mk}[l],
\end{eqnarray}
where $\mu_m$ represents the AGC gain which scales the input to the dynamic range of the ADC at antenna $m$ and $y_m^q[n]$ is the quantized received signal. Let $P_{rx}^m$ denote the average received power at antenna $m$ defined by
\begin{equation}\label{eq:AvePower}
P_{rx}^m = \mathbb{E}[|y_m[n]|^2] = \sum\limits_{k=1}^{K}\beta_k P_k + N_0.
\end{equation}
Note that the average received power is identical for each antenna and $P_{rx}^m = P_{rx}$ for all $m$. The expected value of the AGC gain is the reciprocal of the average received power, $\mu_m = 1/P_{rx}$.
The noise and quantization distortion terms after correlation are 
\begin{eqnarray}
q'_{mk}[l] &=& \frac{1}{\sqrt{N_p}} \sum\limits_{n=0}^{N_p - 1} q_m[n]\phi^*_i[n+l], \\
z'_{mk}[l] &=& \sqrt{\frac{\mu_m}{N_p}} \sum\limits_{n=0}^{N_p - 1} z_m[n]\phi^*_i[n+l]. 
\end{eqnarray}
The linear minimum mean square error (LMMSE) estimator of $\hat{h}_{mk}[l]$ based on $r_{mk}[l]$ is given by 
\begin{equation}
\hat{h}_{mk}[l] = \frac{\sqrt{\beta_kP_kN_p\mu_m}\sigma^2_k[l]}{\beta_kP_kN_p\mu_m \sigma^2_k[l] + E + \mu_m N_0}r_{mk}[l]
\end{equation} 
and the estimation error $\epsilon_{mk}[l] = h_{mk}[l] - \hat{h}_{mk}[l]$ has variance 
\begin{equation}
\mathbb{E} [|\epsilon_{mk}[l]|^2] = (1 - d_k[l])\sigma^2_k[l]
\end{equation}
where 
\begin{equation}
	d_k[l] = \frac{\beta_kP_kN_p\mu_m\sigma^2_k[l]}{\beta_kP_kN_p \mu_m\sigma^2_k[l] + E + \mu_mN_0}. 
\end{equation} 
The estimation error in the frequency domain, $\sfm{e}_{mk}[v] = \sfm{h}_{mk}[v] - \sfm{\hat{h}}_{mk}[v]$ has the variance $1-c_k$ where $c_k$ represents the channel estimation quality and is given by 
\begin{equation}\label{eq:channelQualityIndicator}
c_k = \sum\limits_{l=0}^{L-1} d_k[l]\sigma_k^2[l]. 
\end{equation} 

The estimated channel is utilized for transmission of $N_d = N - N_p$ data symbols. For user $k$, the processed received signal $\hat{x}_k[n]$ is utilized to obtain the transmitted signal $x_k[n]$, where
\begin{equation}
\hat{x}_k[n] = \sum\limits_{m=1}^{M}\sum\limits_{l=0}^{N_d-1} w_{km}[l] y^q_m[n-l]_{N_d},
\end{equation}
$[n]_{N_d} = n \, \text{mod} \, N_d$, and $w_{km}[l]$ is the impulse response of the combining FIR filter with the transfer function $\sfm{w}_{km}[v]$ which is defined by 
\begin{equation}
w_{km}[l] = \frac{1}{N_d} \sum\limits_{v=0}^{N_d-1} \sfm{w}_{km}[v]e^{j2\pi vl/N_d}.
\end{equation}
The processed signal in the frequency domain is 
\begin{equation}\label{eq:EstimatedSignalFreq-1}
\sfm{\hat{x}}_k[v] = \sum\limits_{m=1}^M \sfm{w}_{km}[v]\sfm{y}^q_m[v].
\end{equation}

Given $\sfm{\hat{x}}_k[v]$, the capacity for user $k$ is
\begin{equation} \label{eq:capacity-1}
C = \max_{f_X:~\mathbb{E}[|\sfm{x}_k[v]|^2]\leq 1}  I\left(\sfm{\hat{x}}_k[v];\sfm{x}_k[v]\right)
\end{equation}
where $f_X$ is the distribution of the transmit signal $\sfm{x}_k[v]$. If the transmit signals are assumed to be Gaussian, a lower bound on the capacity can be obtained as \cite{medard2000channelCapacity}
\begin{eqnarray}\label{eq:rateLowerBound}
C \geq R_k \triangleq \log_2 \left(1 + \frac{|\mathbb{E}[\sfm{x}_k^*[v]\sfm{\hat{x}}_k[v]]|^2}{\mathbb{E}[|\sfm{\hat{x}}_k[v]|^2]-|\mathbb{E}[\hat{\sfm{x}_k^*[v]}\sfm{\hat{x}}_k[v]]|^2}\right)
\end{eqnarray}
in bit/s/Hz, where the expectation is taken with respect to symbols and small-scale fading realizations.


In order to compute \eqref{eq:rateLowerBound}, $\sfm{\hat{x}}_k[v]$ can be expanded as
\begin{eqnarray} \label{eq:estimateX-1}
\sfm{\hat{x}}_k[v] &=& \sqrt{\beta_kP_k}\sum\limits_{m=1}^M \sqrt{\mu_m}\sfm{w}_{km}[v] \hat{\sfm{h}}_{mk}[v]\sfm{x}_k[v]  \nonumber \\
 &+& \sqrt{\beta_kP_k}\sum\limits_{m=1}^M \sqrt{\mu_m}\sfm{w}_{km}[v] \sfm{e}_{mk}[v]\sfm{x}_k[v]  \nonumber \\
 &+& \sum\limits_{k'\neq k}^K\sqrt{\beta_{k'}P_{k'}}\sum\limits_{m=1}^M\sqrt{\mu_m}\sfm{w}_{km}[v]\sfm{h}_{mk'}[v]\sfm{x}_{k'}[v]\nonumber \\
 &+& \sum\limits_{m=1}^M \sqrt{\mu_m}\sfm{w}_{km}[v]\sfm{z}_m[v] \nonumber \\
 &+& \sum\limits_{m=1}^M \sfm{w}_{km}[v]\sfm{q}_m[v] 
\end{eqnarray}
where $\sfm{q}_m[v]$ denotes the Fourier transform of the quantization distortion at antenna $m$. Combining \eqref{eq:rateLowerBound} with \eqref{eq:estimateX-1} gives the following lower bound on the capacity of user $k$ using maximum ratio combining (MRC)
\begin{equation}\label{eq:MRC_Capacity-1}
R_k(B_w,M,b) = B_w \left(\frac{N_d}{N}\right) \log_2 \left(1 + \gamma_k\right)~~\text{[bit/s]}
\end{equation}
where we have also taken the bandwidth and pilot overhead into account. The signal-to-interference-noise-and-quantization ratio (SINQR) $\gamma_k$ is given by 
\begin{equation}\label{eq:SINQR}
\gamma_k =  \frac{c_k\beta_kP_kM}{\sum\limits_{k'=1}^K\beta_{k'}P_{k'} + N_0 + P_{rx}E}.
\end{equation}

\section{Problem Statement and Rate Optimization}\label{sec:ProblemStatement_RateOptimization}
We consider the problem of finding the optimum values for the bandwidth, the number of quantization bits and the number of BS antennas to maximize the sum of achievable rates, subject to a fronthaul capacity constraint. Hence, the problem can be stated as
 \begin{equation} \label{RateMaxProblemStatement}
\begin{aligned}
& \underset{B_w,M,b}{\text{maximize}}
& & \sum_{k = 1}^{K} R_k(B_w,M,b) \\
& \text{subject to}
& & B_wMb \leq C_f, \\
\end{aligned}
\end{equation}
 
Here, $M,~b \in \mathbb{Z}_+$ and $B_w \in \mathbb{R}_+$ as any other choice is practically meaningless and $C_f$ is the maximum fronthaul capacity. Note that, the transmission powers are not considered as parameter herein. To find the optimum solution ($B_w^*,M^*,b^*$), we need to investigate the interplay between the variables. Although the subsequent analyses can be applied along with any power control, statistical channel inversion power control is assumed in order to simplify notation \cite{emil2016MIMOmeetsSC}, i.e., user $k$ transmits with 
\beq \label{eq:powerControl-channelinversion}
     P_k = \frac{P}{B_w \beta_k}.
\eeq  
The channel inversion power control results in identical achievable rate for each user, which allows us to denote $R_k = R$ for all $k$ as with \eqref{eq:powerControl-channelinversion} the channel estimation quality and SINQR of the users becomes identical; $c_k = c$ and $\gamma_k = \gamma$ for all $k$ in the subsequent analysis. This assumption reduces the sum rate optimization in \eqref{RateMaxProblemStatement} to the maximization of a single rate expression which applies to all users.

\subsection{Optimization with Fixed Bandwidth}\label{subSec:FixedBandwidth}
First, the interaction between $M$ and $b$ is analyzed for a given $B_w$. Note that for a given $B_w$, the problem reduces to 
\begin{equation} \label{pr:ReducedBwFixed}
\begin{aligned}
& \underset{M,b}{\text{maximize}}
& &  \gamma = \frac{cMP/B_w}{\left(KP/B_w + N_0 \right) \left(1+E\right)} \\
& \text{subject to}
& & Mb \leq C_f/B_w. \\
\end{aligned}
\end{equation}

Our first observation is that the maximum value of $R(B_w,M,b)$ defined in \eqref{eq:MRC_Capacity-1} is attained when the fronthaul capacity constraint is satisfied with equality $Mb = C_f/B_w$ where the integer constraints on $M$ and $b$ are relaxed. This is due to the observation that $R(B_w,M,b)$ is an increasing function with respect to its parameters. Consider the difference between $\gamma(M,b) - \gamma(\bar{M},b+1)$ given by
\beq\label{eq:b_vsM-1}
\frac{c(b)MP/B_w}{\left(KP/B_w + N_0 \right) \left(1+E\right)} - \frac{c(b+1)\bar{M}P/B_w}{\left(KP/B_w + N_0 \right) \left(1+E/4\right)} \gtrless 0,
\eeq
where $\bar{M} = Mb/(b+1)$. When the difference in \eqref{eq:b_vsM-1} is positive, increasing $b$ results in a lower $R(B_w,M,b)$.  
Note that, the channel estimation quality is a function of $b$ and is independent of $M$. Hence, using a lower resolution for quantization decreases the performance of the system through both channel estimation and data transmission processes. In order to identify the regions in which $\gamma$ is decreasing or increasing with respect to $b$ and $M$ based on \eqref{eq:b_vsM-1}, it is sufficient to examine 
\beq\label{eq:b_vsM-2}
\frac{c(b)}{\left(KP/B_w + N_0 \right) \left(1+E\right)} - \frac{\alpha c(b+1)}{\left(KP/B_w + N_0 \right) \left(1+E/4\right)} \gtrless 0
\eeq
where $\alpha = b/(b+1)$. Assuming a uniform power delay profile, $\sigma^2_k[l] = 1/L$ for all $l$, and combining \eqref{eq:channelQualityIndicator} with \eqref{eq:b_vsM-2} we obtain, 
\beq\label{eq:b_vsM-3}
\frac{\left(1+E/4\right)}{ \left(\theta KP/B_w + N_0 + P_{rx}E\right)  } - \frac{\alpha \left(1+E\right)}{\left(\theta KP/B_w + N_0 + P_{rx}E/4\right) } \gtrless 0
\eeq 
where $\theta = N_p/KL$ is called the pilot excess factor \cite{chris2016uplinkOnebit}. \eqref{eq:b_vsM-3} can further be simplified as  
\begin{flalign}\label{eq:b_vsM-4}
\mathsf{I}(\theta-1)&\left(\left(1+E/4\right) -\alpha\left(1+E\right)    \right) +   \notag \\ \left(N_0 + \mathsf{I}\right)&\left(\left(1+E/4\right)^2 -\alpha\left(1+E\right)^2   \right) \gtrless 0
\end{flalign} 
where $\mathsf{I} = KP/B_w$ denotes the total interference which is identical for all users under channel inversion power control and $E$ is a function of $b$ defined in \eqref{eq:QuantizationMSE}. The second term of the summation in \eqref{eq:b_vsM-4} is positive for all values of $b$ and the first term is always positive for $\theta \geq 1$ which allow us to state the following lemma. 

\begin{lemma} \label{lem:Mvsb}
	Consider the uplink system with $\theta \geq 1$ and a given $B_w$, then $b^* = 1$ and $M^* = C_f/B_w$.  
\end{lemma}

\begin{figure}[tb]
	\begin{center}
		\includegraphics[trim=.5cm 0.9cm 0.4cm 1.1cm,clip=true,scale = 0.5]{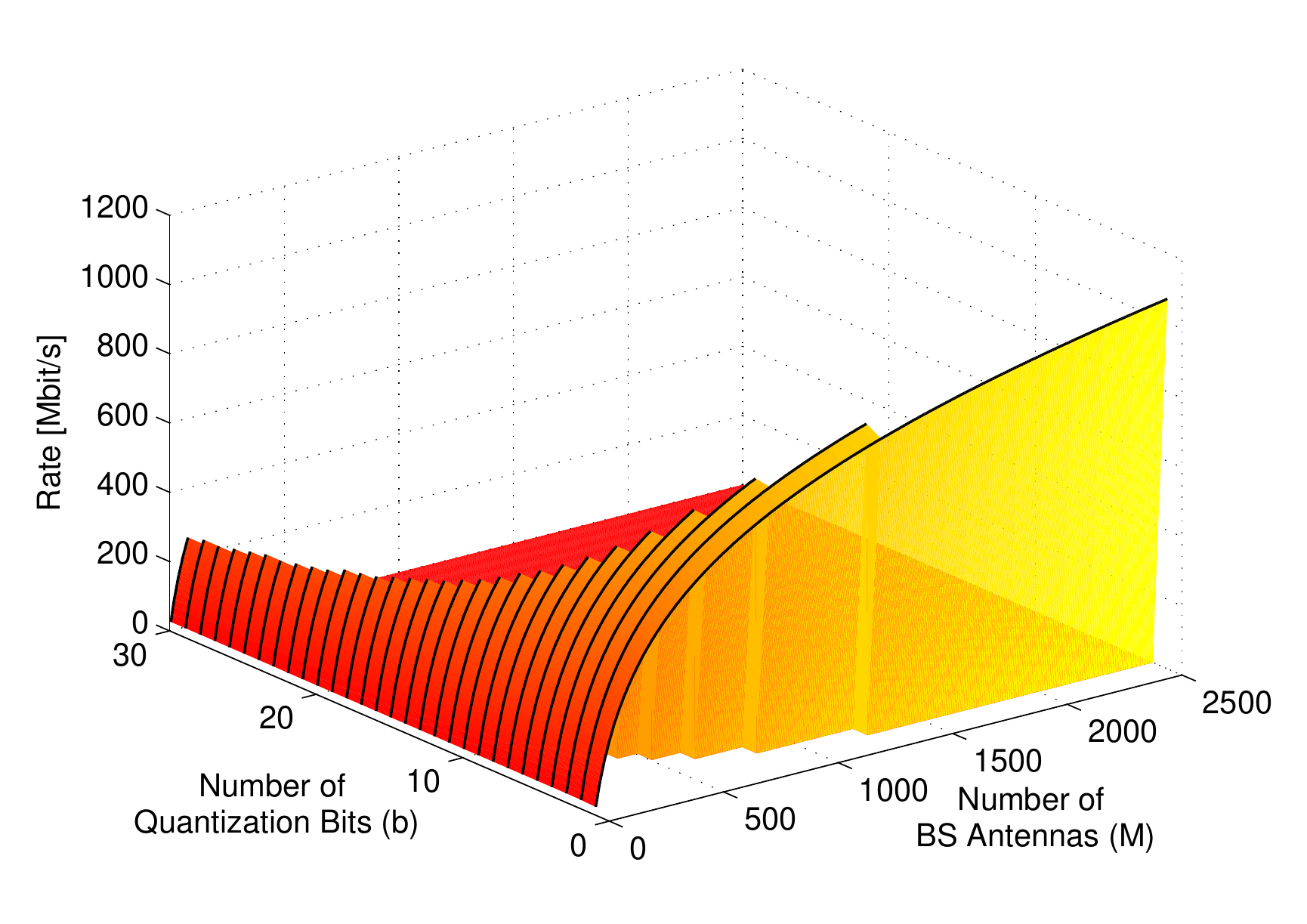}
		\caption{Achievable rate as a function of $M$ and $b$ for $K = 20$. }
		\label{fig:figMvsb}
	\end{center}{\vspace{-7mm}}
\end{figure}

 In Fig. \ref{fig:figMvsb}, a system with a bandwidth of $B_w = 200$ MHz, $20$ users, $15$ dB SNR and a $500$ Gbit/s fronthaul capacity is considered and the achievable rate as function of $M$ and $b$ is depicted. Note that $R(M,b)$, the achievable rate as a function of $M$ and $b$, is an increasing function with respect to its parameters, i.e., without any constraints increasing either results in a higher rate value. However, for the case with a fronthaul capacity constraint, maximizing $R(M,b)$ requires using $1$-bit ADCs with maximum number of $M$. 
 In other words, for an allocated $Mb$ bits to describe the received signal, it is better to receive bits from a higher number of different antennas instead of a better description from a smaller number of antennas. 
 
\subsection{Optimization with Fixed Number of Antennas}\label{subSec:Fixed_M}
\begin{figure}[tb]
	\begin{center}
		\includegraphics[trim=.7cm .3cm 0cm .3cm,clip=true,scale = 0.5]{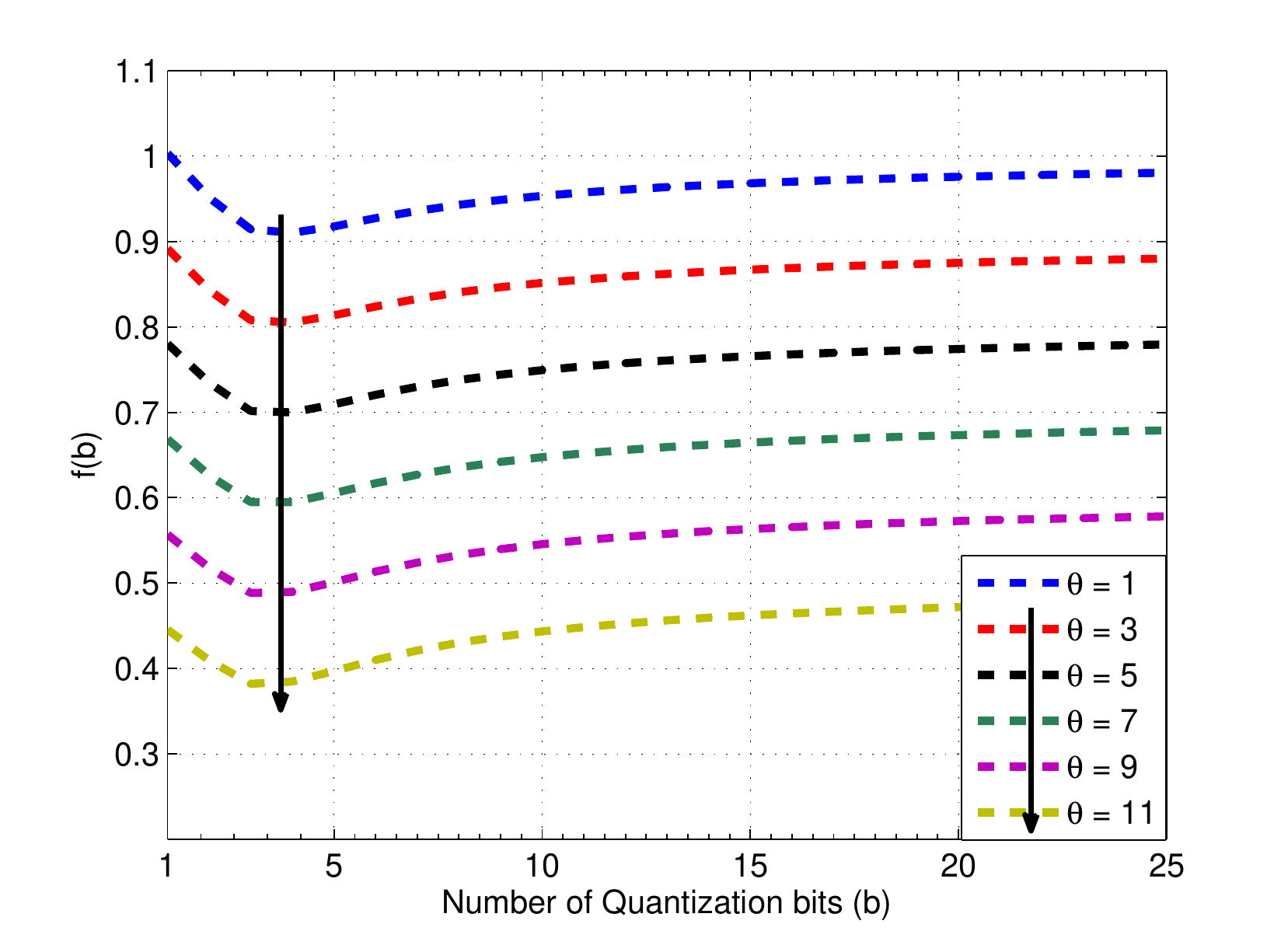}{\vspace{-1mm}}
		\caption{Threshold function with respect to the number of quantization bits.}{\vspace{-6mm}}
		\label{fig:fig_bvsfb}
	\end{center}
\end{figure}

Next consider the change of rate with respect to $B_w$ and $b$ for a given $M$. In this case, the examination of SINQR values is not sufficient since $B_w$ affects the pre-log factor and the change on \eqref{eq:MRC_Capacity-1} must be considered as follows:
\begin{flalign}
B_w \left(\frac{N_d}{N}\right) \log_2\left(1 + \gamma(b,B_w)\right) -\notag \\ \bar{B}_w \left(\frac{N_d}{N}\right) \log_2\left(1 + \gamma(b+1,\bar{B}_w)\right) \gtrless 0 \label{eq:BwVsb-1}
\end{flalign}    
where $\bar{B}_w = \alpha B_w$. The regions where \eqref{eq:BwVsb-1} is negative corresponds to the regions where increasing $b$ results in a higher rate value. \eqref{eq:BwVsb-1} can be simplified as

\begin{flalign}
\frac{1}{\alpha}\log_2\left(1 + \gamma(b,B_w)\right) - \log_2\left(1 + \gamma(b+1,\bar{B}_w)\right)& \gtrless 0 \notag \\
\log_2\left(\left(1 + \frac{P^2MN_p/L}{ \left(KP + N_0B_w\right)^2 \left(1+E\right)^2   }\right)^{1/\alpha}\right) - \notag \\ \log_2\left(1 + \frac{P^2MN_p/L}{ \left(KP + \alpha N_0B_w\right)^2 \left(1+E/4\right)^2   }\right) \gtrless 0 \label{eq:BwVsb-2}
\end{flalign}    
where we assumed $\theta = 1$ ($N_p = KL$) for mathematical tractability. Utilizing the monotonicity of logarithmic functions and Bernoulli's inequality, we obtain
\begin{flalign}
\frac{1}{\alpha \left(KP + N_0B_w\right)^2 \left(1+E\right)^2 } -& \notag \\ \frac{1}{\left(KP + \alpha N_0B_w\right)^2 \left(1+E/4\right)^2} >&~ 0 \label{eq:BwVsb-4}
\end{flalign}    
Rearranging the terms leads to 
\begin{flalign}
\frac{\left(KP + \alpha N_0B_w\right) \left(1+E/4\right) - \sqrt{\alpha} \left(KP + N_0B_w\right) \left(1+E\right)}{\alpha \left(KP + N_0B_w\right)^2 \left(1+E\right)^2\left(KP + \alpha N_0B_w\right)^2 \left(1+E/4\right)^2} > 0 \notag 
\end{flalign} 
 This allows us to identify the regions in which increasing $B_w$ results in a higher rate:
\begin{equation} \label{eq:BwVsb-Th}
\frac{KP}{B_wN_0} > f(b)
\end{equation}
where 
\begin{equation}
f(b) = \alpha\frac{ \left(-1 - E/4 + 1/\sqrt{\alpha} + E/\sqrt{\alpha}\right)}{1 + E/4 - \sqrt{\alpha} - E\sqrt{\alpha}}
\end{equation}
is the threshold function that only depends on $b$. It is straight-forward to show that both the numerator and denominator of $f(b)$ are positive for finite values of $b$. Fig. \ref{fig:fig_bvsfb} depicts the change of $f(b)$ with respect to $b$. An important point is that, except for the case where $b = 1$, $f(b) < 1$ and this implies that increasing $B_w$ when the interference is greater than the noise power results in a higher rate. Furthermore, increasing $\theta$ results in a lower threshold value as illustrated in Fig. \ref{fig:fig_bvsfb}. For $\theta > 1$, the threshold is no longer only a function of $b$ and decreases with increasing $\theta$ which favors utilizing larger $B_w$.  
 
 Note that the condition provided by \eqref{eq:BwVsb-Th} is a sufficient condition due to the approximation in \eqref{eq:BwVsb-4} via Bernoulli's inequality. Hence, increasing the number of bits in the regions where \eqref{eq:BwVsb-Th} is not satisfied, does not necessarily provide a better performance.  
The analysis on the interplay between $B_w$ and $b$ for a given $M$ allows us to state the following:    
\begin{lemma}\label{lem:BwVsb}
	Assume that \eqref{eq:BwVsb-Th} holds for a given $M$, then $b^* = 1$ and $B_w^* = C_f/M$. 
\end{lemma}

\begin{figure}[tb]
	\begin{center} 
		\includegraphics[trim=.4cm .1cm .4cm 0.6cm,clip=true,scale = 0.5]{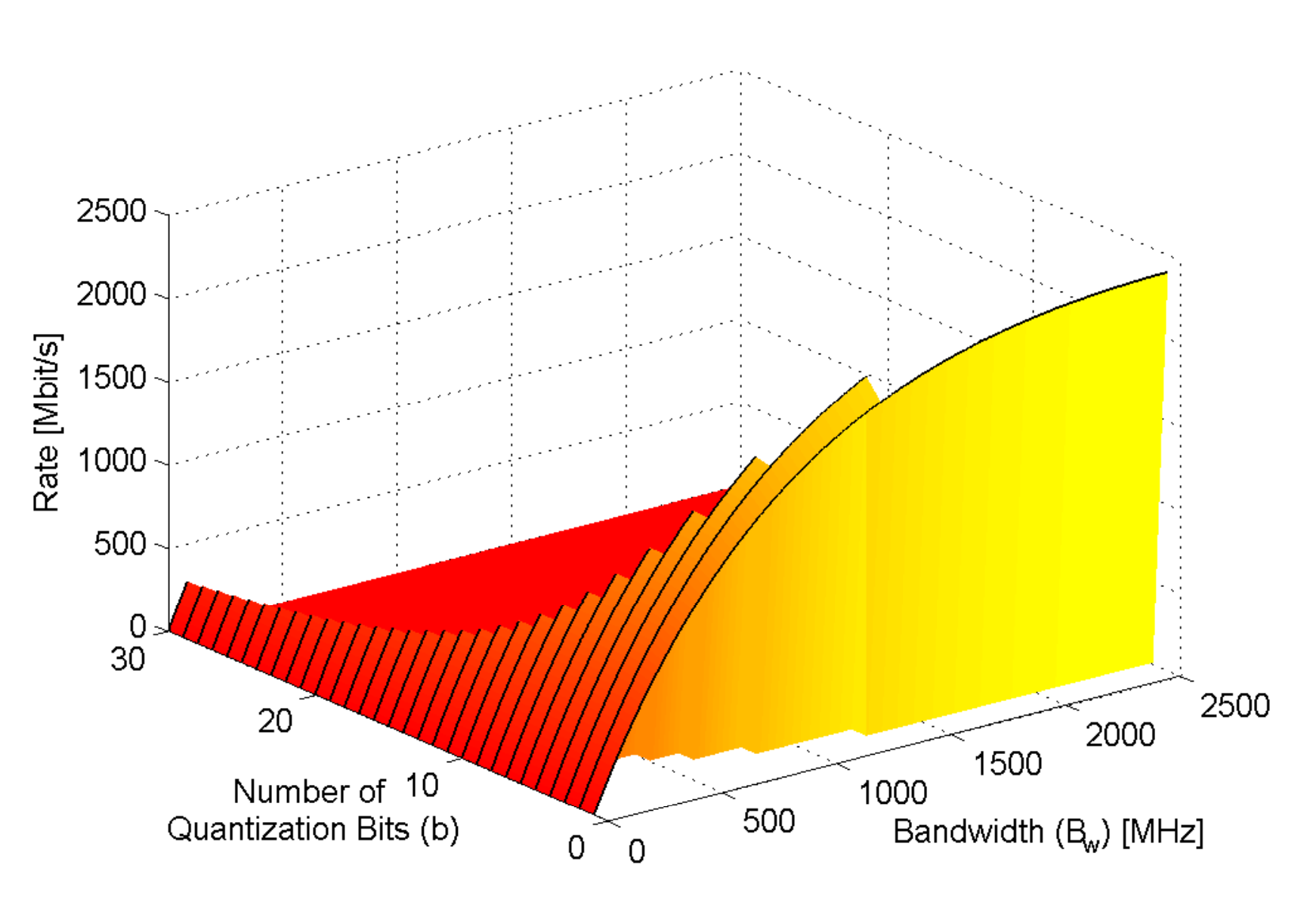}{\vspace{-4mm}}
		\caption{Achievable rate as a function of $B_w$ and $b$ for $K = 20$. }
		\label{fig:figBwvsb}
	\end{center} {\vspace{-5mm}}
\end{figure}

In Fig. \ref{fig:figBwvsb}, the change on rate with respect to the bandwidth and number of quantization bits is demonstrated. For this particular example, $M=200$, $K=20$ and $C_f = 500$ Gbit/s with $15$ dB SNR. The maximum rate value is achieved when $b = 1$ and $B_w = C_f/M$.

\begin{theorem} \label{thm:b=1}
 Assume $\theta \geq 1$ and \eqref{eq:BwVsb-Th} holds. Let the triplet $(B_w^*, M^*, b^*)$ denote the optimal solution to the problem in \eqref{RateMaxProblemStatement}. Then, $b^* = 1$ for any $(B_w^*, M^*)$ pairs. 
\end{theorem}
\begin{IEEEproof}
Assume $b^* > 1$ and consider a setup with $b=1$, then based on Lemma \ref{lem:Mvsb} and Lemma \ref{lem:BwVsb} a higher rate value can be achieved with choosing $b = 1$ and using a higher $M$ and/or $B_w$ which contradicts with the assumption $b^* > 1$. 
\end{IEEEproof}

\subsection{Optimization with Fixed Number of Quantization Bits}\label{subSec:Fixed_b}
In the final part of the analysis, we investigate the interaction between $M$ and $B_w$ for a given $b$. The approaches used to analyze the interaction between $b$ and $M$, $B_w$ leads to inconclusive results which compels a different analysis technique. Since the maximum rate is achieved when the fronthaul capacity constraint is satisfied with equality, it is sufficient to consider the curve defined by $M\cdot B_w = C_f$. Furthermore, we can reduce the dimension of the search space by introducing the auxiliary variable $s \in (1/C_f,1]$ and defining $\bar{M} = 1/s$, $\bar{B}_w = C_fs$. Note that, $\bar{M} \cdot \bar{B}_w = C_f$ for any value of $s$. Let $\upsilon = N_dC_f/(N\ln2)$, then \eqref{eq:MRC_Capacity-1} can be re-written as 
\begin{flalign}
R(s) =  \upsilon s \ln \left(1 + \frac{c P/C_fs^2}{KP/C_fs + N_0 + P_{rx}E}\right)
\end{flalign} 
which can be combined with \eqref{eq:channelQualityIndicator} to obtain
\begin{flalign}\label{eq:MvsBw-1}
R(s) =  \upsilon s \ln \left(1 + \frac{  P^2N_p/Ls}{ \tau(\theta, s)+\left(1+E\right)^2\left(KP + C_fsN_0\right)^2}\right)
\end{flalign}   
where 
\begin{equation}\label{eq:tau}
\tau(\theta, s) = (\theta - 1)KP\left(KP + C_fsN_0\right)\left(1+E\right).
\end{equation}
It is straightforward to show that \eqref{eq:MvsBw-1} is a concave function of $s$ by showing the second derivative is non-positive. 
Hence, the maximizing value 
$s^*$ can be obtained by equating to zero. The derivative of \eqref{eq:MvsBw-1} is  
\begin{equation}\label{eq:MvsBw-2}
\frac{dR(s)}{ds} =  \upsilon\ln \left(1+\omega\right) +\upsilon s \frac{\dot{\omega}}{1+\omega} 
\end{equation} 
where $\dot{w}$ denotes the derivative of $\omega$ with respect to $s$ and 
\begin{equation}\label{eq:w}
w =  \frac{P^2N_p/Ls}{\tau(\theta, s) +\left(1+E\right)^2\left(KP + C_fsN_0\right)^2}.
\end{equation}  
Although, \eqref{eq:MvsBw-2} can easily be computed numerically, it is a challenging task to find an analytical expression for $s^*$. Furthermore, it is difficult to understand the behavior of \eqref{eq:MvsBw-2} with respect to $B_w$ and $M$. Fortunately, the interplay between the variables can be understood by investigating the sign of the derivative with respect to $s$ as it determines the regions where increasing $s$ results in a higher rate value and vice versa.
 To this end, Pad\'e approximants \cite{topsok2006logBounds} are utilized to approximate the logarithmic term which results in 
\begin{equation}\label{eq:MvsBw-Result-1}
\frac{KP}{B_wN_o} > 4\left(1+E\right)^2 \left(KP + B_wN_0\right) \frac{L}{MP^2N_p} + 1
\end{equation} 
and allows us to make some general observations regarding $M$ and $B_w$. First, it is harder to satisfy \eqref{eq:MvsBw-Result-1} at higher $B_w$ values because it requires either a larger $N_p$ value which corresponds to better channel estimates, a higher $P/N_0$ value which can be considered as  the effective SNR and/or increased number of antennas. Second, employing ADCs with more quantization bits favors utilization of larger $B_w$. In short, the users who have good signal quality will benefit more from utilizing larger bandwidth instead of more antennas. 

Next, the following upper bound is used on the logarithmic term in \eqref{eq:MvsBw-2}  \cite{topsok2006logBounds}
\begin{equation}
\ln(1+x) \leq \frac{x}{2}\cdot \frac{2+x}{1+x},~~\text{for}~0\leq x\leq \infty 
\end{equation} 
which leads to  
\begin{equation}\label{eq:MvsBw-result-2}
M < 4(1+E)^2\left(KP + B_wN_0\right) \frac{B_wN_0L}{P^2N_p}.
\end{equation}
Hence, while \eqref{eq:MvsBw-result-2} is satisfied, increasing $M$ while reducing $B_w$ provides a higher rate. As $M$ increases satisfying \eqref{eq:MvsBw-result-2} requires a larger value of $B_w$. Contrary to the threshold for $B_w$ given in \eqref{eq:MvsBw-Result-1}, having a low SNR, low-bit ADCs and shorter pilot sequences favor increasing $M$. By comparing the thresholds defined in \eqref{eq:MvsBw-Result-1} and \eqref{eq:MvsBw-result-2}, it can be concluded that having good channel estimates, high number of quantization bits and high SNR favors increasing $B_w$ whereas a system with low SNR and small number of quantization bits benefits most from having more antennas. 


  \begin{figure}[thb]
  	\begin{center}{\vspace{-2mm}}
  		  	\includegraphics[trim=.3cm .7cm .1cm 0.9cm,clip=true,scale = 0.5]{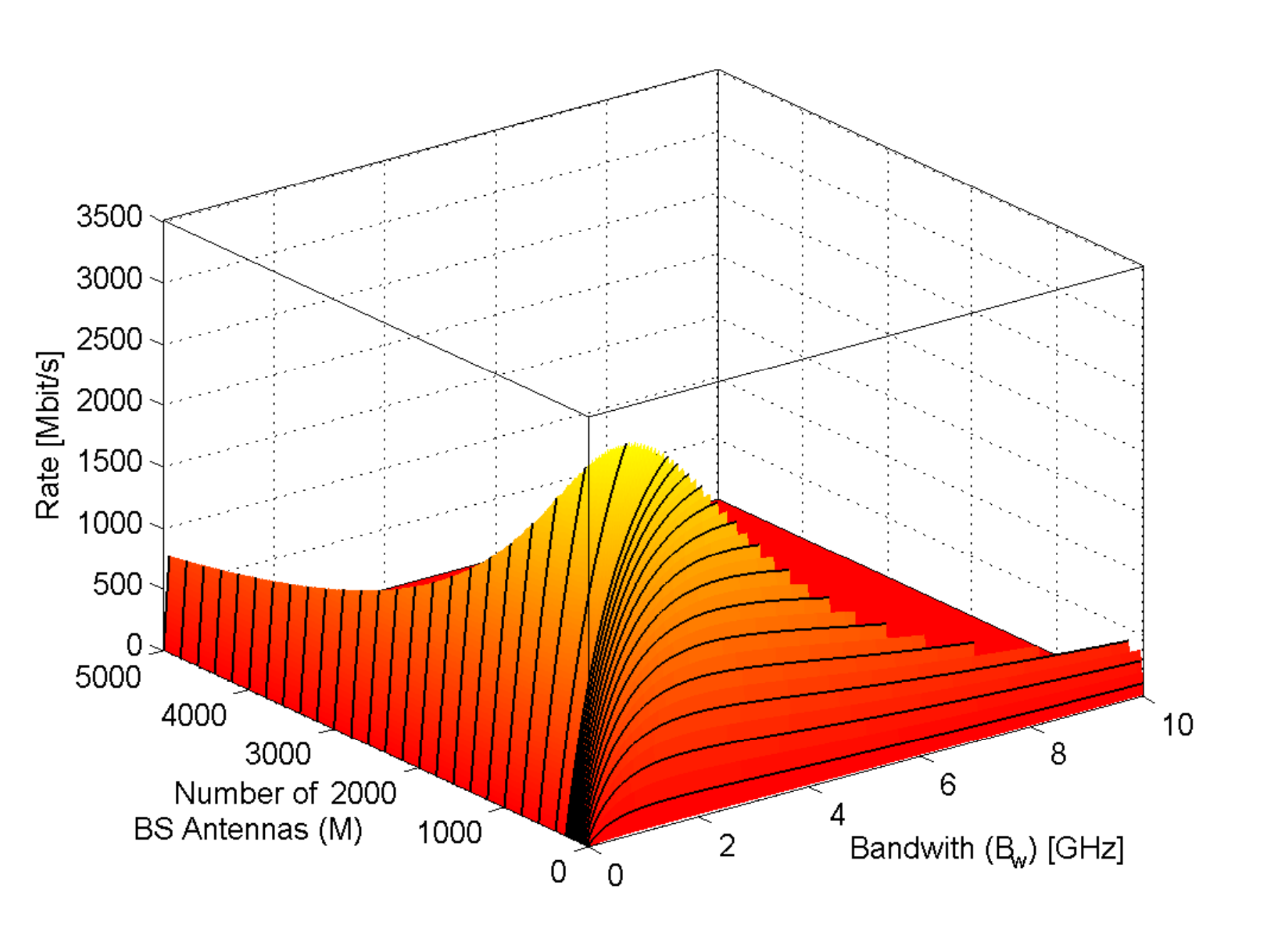}
  		\caption{Achievable rate as a function of $M$ and $B_w$ for a setup with $20$ users and $1$-bit ADCs. Maximum rate is achieved at $M^* = 381$ and $B_w^* = 131$ MHz for this particular example.  }
  		\label{fig:figMvsBw-killer}
  	\end{center}{\vspace{-5mm}}
  \end{figure}
       
Fig. \ref{fig:figMvsBw-killer} illustrates the achievable rate for various ($M$, $B_w$) pairs in a setup with $15$ dB SNR, $20$ users and $1$-bit ADCs. Contrary to the previous cases with the quantization bits, the interplay between $M$ and $B_w$ is harder to interpret and the optimum value depends on various parameters.

\section{Numerical Analysis} \label{sec:Numerical}

 A single centralized BS with a $350\,$m radius and $20$ underlying, uniformly distributed users are considered. The simulation parameters are mainly acquired from \cite{emil2016MIMOmeetsSC} and the references therein. The path loss model is given as $\beta_k = -130 - 37.6\log_{10}\left( d_k\right)\,$dB  where $d_k$ denotes the distance of user $k$ to the BS in kilometers. 

The transmission powers of users are given by \eqref{eq:powerControl-channelinversion} which results in identical received power at the BS. The user with the worst channel condition  transmits with maximum power, i.e., $P = P_{max}\beta_{min}$ where $\beta_{min} = \min_{k} \beta_k$. Hence user $k$ transmits with 
\begin{equation}\label{eq:PowerUserk}
P_k = \frac{P_{max}\beta_{min}}{B_w\beta_k}.
\end{equation} 
Since we will change the bandwidth, instead of the usual SNR definition, we use a reference SNR, $\Gamma = P_{max}\beta_{edge}/(10^6 N_0)$, as a parameter which represents the SNR value of a user at the cell edge with $1$ MHz, where $\beta_{edge}$ is the large scale fading of a user at cell edge. 

\begin{figure}[thb]
 	\begin{center}
 		\includegraphics[trim=.8cm .1cm .4cm .6cm,clip=true,scale = 0.55]{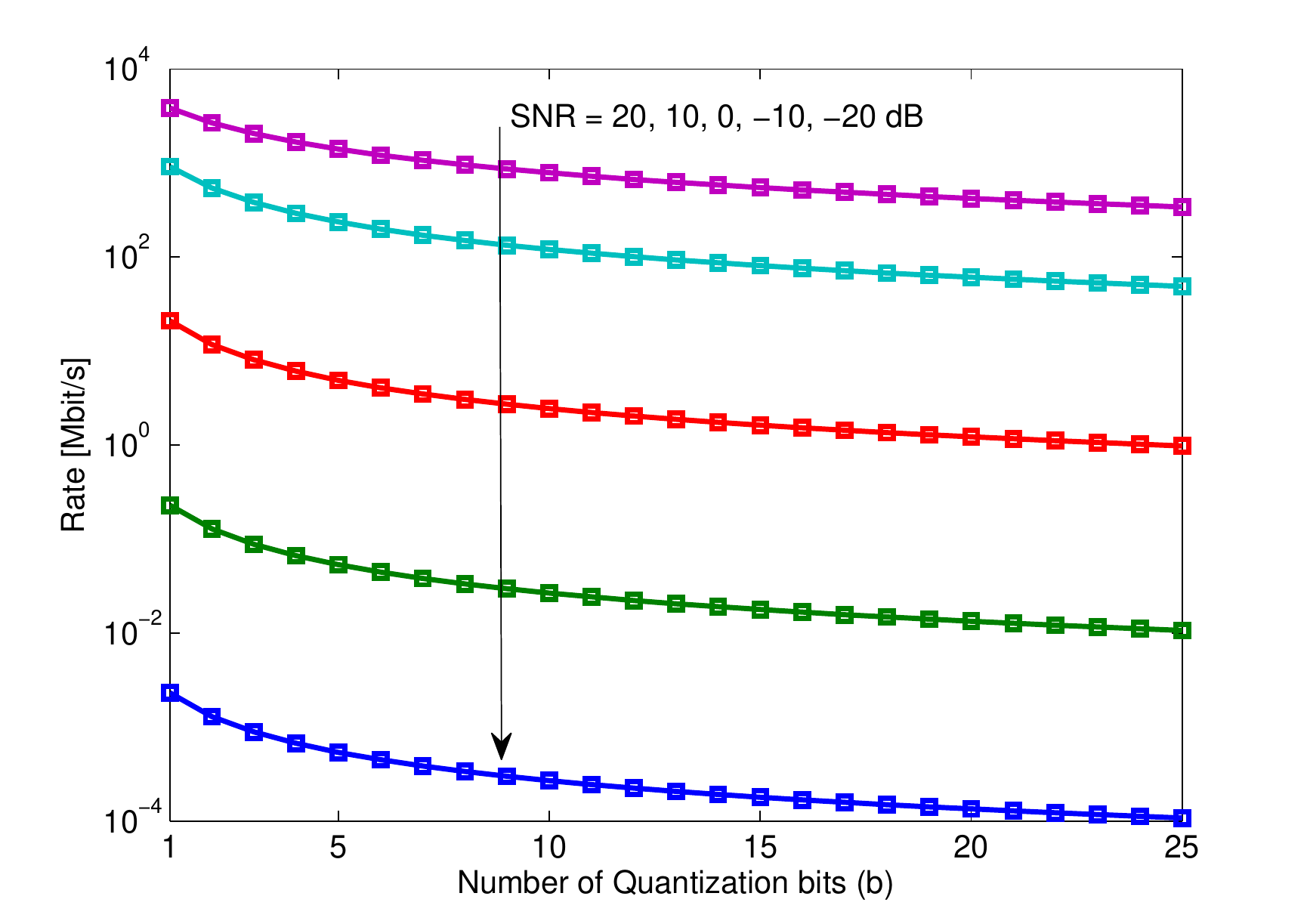}{\vspace{-2mm}}
 		\caption{Achievable rate as a function of $b$ and $M$ for various SNR values. }{\vspace{-1mm}}
 		\label{fig:figMvsb-SNR-v2}
 	\end{center}
 \end{figure}

In the first example, the rate is considered as a function of $b$ and $M$ to verify the result of Lemma \ref{lem:Mvsb}. Fig. \ref{fig:figMvsb-SNR-v2} depicts the achievable rate for a fixed $B_w = 200$ MHz and $C_f = 500$ Gbit/s. As the maximum rate is achieved when the fronthaul capacity constraint is satisfied with equality, the number of antennas is $C_f/(B_w\cdot b)$ for any $b$ value.  
 $1000$ Monte Carlo trials are utilized to obtain the achievable rate at each SNR value. The simulations reveal that the maximum rate is attained when $b = 1$ at each SNR value. Furthermore, the behavior of rate with respect to $b$ and $M$ is independent of the SNR value. As expected, higher rate values are achievable with increasing SNR. However, the gains from increasing SNR reduces as the rate function gets closer to the band-limited region.    

\begin{figure}[tb]
	\begin{center}{\vspace{-3mm}}
		\includegraphics[trim=.9cm .3cm 1.2cm .7cm,clip=true,scale = 0.55]{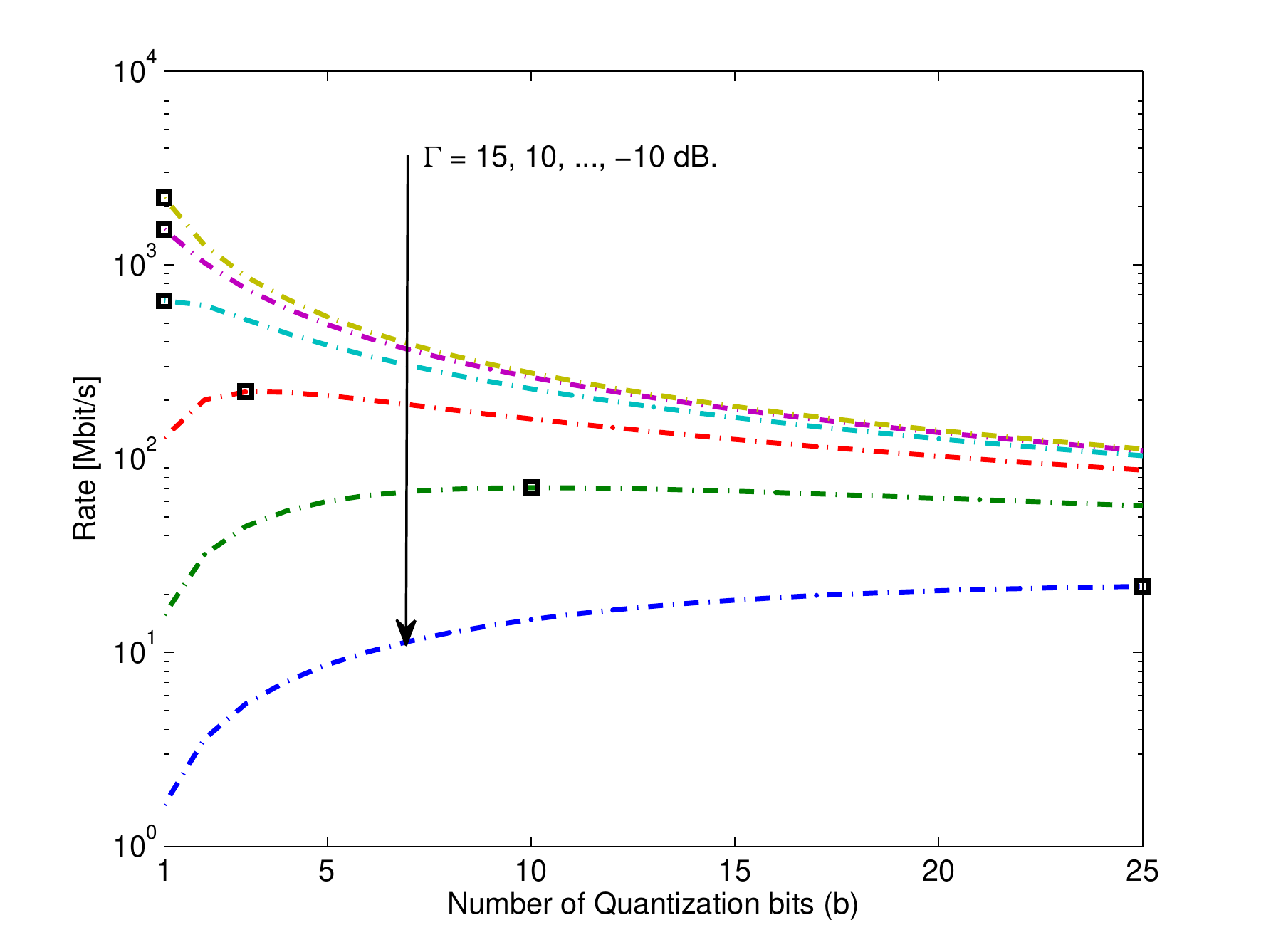}{\vspace{-2mm}}
		\caption{Achievable rate as a function of $b$ and $B_w$ for various SNR values with $1000$ antennas. The squares indicate the maximum value of the curves. }{\vspace{-4mm}}
		\label{fig:figBwvsb-SNR-v1}
	\end{center}
\end{figure}

Next, an example which provides insight into the relation between $B_w$ and $b$ is presented. Fig. \ref{fig:figBwvsb-SNR-v1} depicts the rate as a function of $b$ and $B_w$ for various reference SNR values. For this example, a fixed $M = 1000$ is utilized. The bandwidth is chosen as $B_w = C_f/(M\cdot b)$ at each point, i.e., the initial $B_w$ value (when $b=1$) is $500$ MHz and each curve denotes the rate for a particular reference SNR value. The lower $\Gamma$ values favor increasing the number of quantization bits and as $\Gamma$ increases the maximum rate can be reached with larger $B_w$.

\begin{figure}[tb]
	\begin{center}
		\includegraphics[trim=.8cm .2cm .4cm .6cm,clip=true,scale = 0.6]{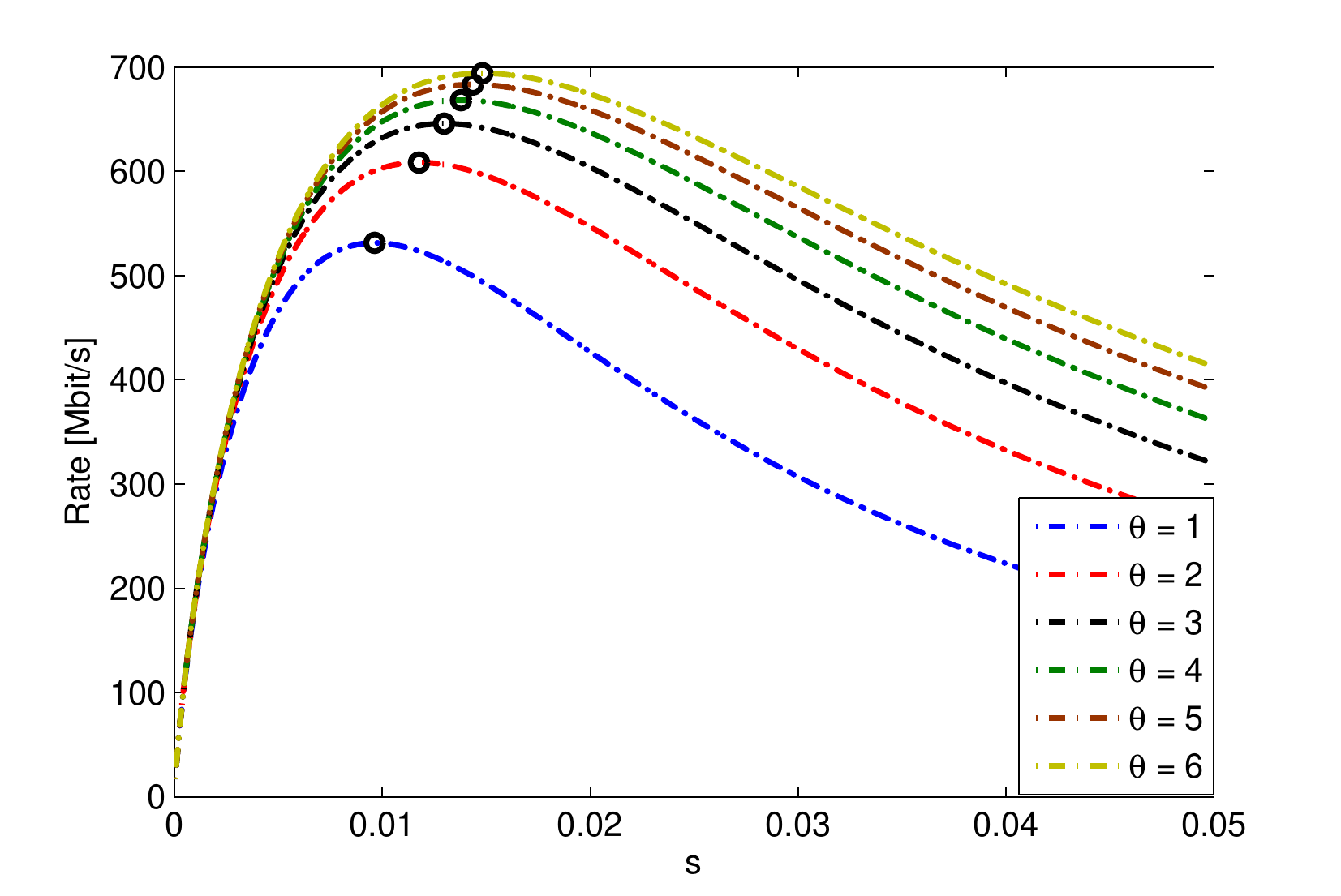}{\vspace{-2mm}}
		\caption{Achievable rate as a function of $s$ for various $\theta$ values with $20$ users. The circles illustrate the maximum value of the curves. }
		\label{fig:figMvsBw-1}{\vspace{-6mm}}
	\end{center}
\end{figure}

Fig. \ref{fig:figMvsBw-1} illustrates the achievable rate with respect to $s$ for various $\theta$ values. Recall that, $s$ is the auxiliary variable that allows us to define $M = 1/s$ and $B_w = C_f s$. Hence, a lower value of $s$ corresponds to a higher $M$ value and vice versa. As expected as $\theta$ increases, a better estimate of the channel can be obtained which leads to higher rate values. Furthermore, the gain is most significant between $\theta = 1$ and $\theta = 2$ which agrees with the analysis presented in \cite{chris2016uplinkMultibit}. Note that, as the channel conditions improve the bandwidth utilized at the maximum rate increases which is in alignment with the theoretical analysis exhibited in Section \ref{sec:ProblemStatement_RateOptimization}.

\section{Conclusion}
This paper has designed a communication system that maximizes the sum rate under constrained fronthaul capacity. The optimization variables were the number of BS antennas $M$, the number of quantization bits $b$ and the bandwidth $B_w$. A simple analytical expression for the achievable rate was derived and utilized to investigate the interplay between the optimization variables. The analytical results reveal that operating at $1$-bit quantization resolution while maximizing the number of antennas is optimum for a given $B_w$, whereas the interplay between $B_w$ and $M$ depends on various parameters, such as SNR. Hence, this nontrivial tradeoff implies that we want a combination of Massive MIMO and large bandwidths in practical systems.




\ifCLASSOPTIONcaptionsoff
  \newpage
\fi


\begin{thebibliography}{1}
\bibitem{qual1000}
``The 1000x data challenge'', Tech. Rep., Qualcomm.
\bibitem{zander2017beyondUDB}
J.~Zander, ``Beyond the Ultra-Dense Barrier: Paradigm Shifts on the Road Beyond 1000x Wireless Capacity'', \emph{IEEE Wireless Commun.}, 2017.
\bibitem{hoydis2011smallcell}
J.~Hoydis, M.~Kobayashi and M.~Debbah, ``Green small-cell networks'', \emph{IEEE Veh. Technol. Mag.}, vol.~6, no.~1, pp.~37-43, 2011.
\bibitem{hwang2013holisticSmallcell}
I.~Hwang, B.~Song and S.~S.~Soliman, ``A holistic view on hyper-dense heterogeneous and small cell networks'', \emph{IEEE Commun. Mag.}, vol.~51, no.~6, pp.~20-27, 2013.
\bibitem{erik2014MIMOmag}
E.~G.~Larsson, O.~Edfors, F.~Tufvesson and T.~L.~Marzetta, ``Massive MIMO for next generation wireless systems'', \emph{IEEE Commun. Mag.}, vol.~52, no.~2, pp.~186--195, 2014.
\bibitem{niu2015mmWave}
Y.~Niu, Y.~Li, D.~Jin, L.~Su and A.~V.~Vasilakos, ``A survey of millimeter wave communications (mmWave) for 5G: opportunities and challenges'', \emph{Wireless Networks}, vol.~21, no.~8, pp.~2657--2676, 2015.
\bibitem{andrews2016limits}
J.~G.~Andrews, X.~Zhang, G.~D.~Durgin and A.~K.~Gupta, ``Are we approaching the fundamental limits of wireless network densification?'', \emph{IEEE Commun. Mag.}, vol.~54, no.~10, pp.~184--190, 2016.
\bibitem{redBook}
T.~L.~Marzetta, E.~G.~Larsson, H.~Yang and H.~Q.~Ngo, ``Fundamentals of Massive MIMO'', \emph{Cambridge University Press}, 2016.
\bibitem{andrews2014whatWill5Gbe}
J.~G.~Andrews, S.~Buzzi, W.~Choi, S.~V.~Hanly, A.~Lozano, A.~C.~K.~Soong and J.~C.~Zhang, ``What will 5G be?'', \emph{IEEE J. Sel. Areas Commun.}, vol.~32, no.~6, pp.~1065--1082, 2014.
\bibitem{5gppp2016RRHwhite}
5G PPP Architecture Group, ``View on 5G Architecture'',
July 2016.
\bibitem{peng2015NFV}
M.~Peng, C.~Wang, V.~Lau, and H.~V.~Poor, ``Fronthaul-constrained cloud radio access networks: Insights and challenges'', \emph{IEEE Wireless Commun.}, vol.~22, no.~2, pp.~152--160, 2015.

\bibitem{gomes2015fronthaul}
N.~J.~Gomes, P.~Chanclou, P.~Turnbull, A.~Magee and V.~Jungnickel, ``Fronthaul evolution: From CPRI to ethernet'', \emph{Optical Fiber Technology}, vol.~26, pp.~50--58, 2015.
\bibitem{chris2016uplinkOnebit}
C.~Moll\'en, J.~Choi, E.~G.~Larsson, and R.~W.~Heath, ``Uplink performance of wideband massive MIMO with one-bit ADCs'', \emph{IEEE Trans. Wireless Commun.}, 2016.

\bibitem{desset2015validation}
C.~Desset and L.~Van der Perre, ``Validation of low-accuracy quantization in massive MIMO and constellation EVM analysis'', \emph{European Conference on Networks and Communications}, pp.~21--25, 2015.
\bibitem{jacobsson2017lowADCmimo}	
S.~Jacobsson, G.~Durisi, M.~Coldrey, U.~Gustavsson and C.~Studer, ``Throughput analysis of massive MIMO uplink with low-resolution ADCs'', \emph{IEEE Trans. Wireless Commun.}, 2017.


%


%


	


\bibitem{sarajlic2016eeMIMOquantization}
M.~Sarajli\'{c}, L.~Liu, and O.~Edfors, ``An Energy Efficiency Perspective on Massive MIMO Quantization'', [Online]. Available:https://arxiv.org/abs/1612.02320

\bibitem{chris2016uplinkMultibit}
C.~Moll\'en, J.~Choi, E.~G.~Larsson, and R.~W.~Heath, (2016) ``Achievable Uplink Rates for Massive MIMO with Coarse Quantization'', [Online] https://arxiv.org/abs/1611.05723 

\bibitem{medard2000channelCapacity}
M.~Medard, ``The effect upon channel capacity in wireless communications of perfect and imperfect knowledge of the channel'', \emph{IEEE Trans. Inf. Theory}, vol.~46, no.~3, pp.~933--946, 2000.

\bibitem{emil2016MIMOmeetsSC}
E.~Bj\"{o}rnson, L.~Sanguinetti, and M.~Kountouris, ``Deploying dense networks for maximal energy efficiency: Small cells meet massive MIMO'', \emph{IEEE J. on Sel. Areas Commun.}, vol.~34, no.~4, pp.~832--847, 2016.




\bibitem{topsok2006logBounds}
F.~Topsok, ``Some bounds for the logarithmic function'', \emph{Inequality theory and applications}, vol.~4, 2006.

\end{thebibliography}

%

%
%
%




\end{document}